\documentclass[twocolumn,showpacs,preprintnumbers,amsmath,amssymb]{revtex4}
\usepackage{graphicx}
\begin{document}

\title{Thermodynamics of vesicle growth and instability}

\author{Duccio Fanelli}
\affiliation{Dipartimento di Energetica, 
Via S. Marta 3, 50139 Florence, Italy
}
\author{Alan J. McKane}
\affiliation{ 
Theoretical Physics, School of Physics and Astronomy,
University of Manchester, Manchester M13 9PL, United Kingdom
} 
 
\begin{abstract}
We describe the growth of vesicles, due to the accretion of lipid molecules to
their surface, in terms of linear irreversible thermodynamics. Our treatment 
differs from those previously put forward by consistently including the energy 
of the membrane in the thermodynamic description. We calculate the critical 
radius at which the spherical vesicle becomes unstable to a change of shape in
terms of the parameters of the model. The analysis is carried out both for the
case when the increase in volume is due to the absorption of water and when a
solute is also absorbed through the walls of the vesicle. 
\end{abstract} 


\maketitle
 
\vspace{0.8cm}

\section{Introduction}
\label{intro}
Vesicles are small cell-like structures in which the membrane separating the 
contents of the vesicle from the environment takes the form of a lipid bilayer.
Part of their appeal comes from the fact that living cells are essentially
very complex vesicles --- with the membrane containing mixtures 
of different lipids and other components, a cytoskeleton and complex surface 
structures\,\cite{alb02}. This has led to the use of vesicles as the basic 
component of models of protocells\,\cite{dea86,mor88,our99}.
On the other hand simple vesicles, without any additional structure, have many 
fascinating properties when observed in the laboratory. Their self-assembly, 
their growth, their shape and the fact that they divide to produce daughter 
vesicles have many aspects which are little-understood. The latter property 
of replication is especially interesting in the context of models of 
protocells. One can ask: how much of simple protocell dynamics can be explained
using the statistical thermodynamics of vesicles, without the introduction of 
more complex processes or of genetic material? This question will be the 
motivation for the present work. In particular we will be interested in 
describing the dynamics of vesicle growth and the instability which leads to 
the vesicle changing shape.

To begin the construction of a model for vesicles, it is first necessary to 
review their dimensions. The typical thickness of the bilayer is $7$nm, 
while the radius of the vesicle itself can be anything from several times 
this value up to $100\,\mu$m for so-called giant vesicles. Therefore the 
bilayer can be thought of as a thin membrane or shell enclosing the contents 
of the vesicle. In the case of biomolecules the standard picture of this 
membrane is the fluid mosaic model\,\cite{sin72} where the proteins, enzymes 
and other such constituents are embedded in the lipid bilayer in which the 
lipid molecules can freely move as in a liquid. The proteins and enzymes may 
be able to move from the inner part of the bilayer to the outer part which is
in contact with the environment. In the case of interest to us here, these 
biological aspects are absent and only the fluid nature of the bilayer 
remains. Therefore the picture which emerges is of a two-dimensional surface, 
which is supplemented with a thin fluid layer on either side to describe the 
physical aspect of the bilayer\,\cite{sei97}. Turning this characterization 
into a quantitative description can be achieved in several different ways; 
the resulting models go under names such as the spontaneous curvature 
model\,\cite{hel73} or the bilayer couple model\,\cite{sve82,sve83,sve89}. 
However these models have the common feature of a bending energy of the 
membrane, which is given in terms of the curvature of the two-dimensional 
surface, together with some extra feature which, in some very basic way, 
accounts for the fact that the membrane has a thickness. In the spontaneous 
curvature model this is a extra factor $C_{0}$ subtracted from twice the mean 
curvature of the surface, and in the bilayer couple model it is a constraint 
on the mean curvature. Here we will use the spontaneous curvature model, for 
which the bending energy is
$E = (\kappa/2) \int dA \left( 2H - C_{0} \right)^{2}$, where $H$ is the local
mean curvature, $A$ is the surface area $A$, and $\kappa$ is the bending 
rigidity.

The vast majority of studies of vesicles based on this form of the bending 
energy have been of a purely static nature; the energy $E$ has been minimized 
subject to the constraints that the area of the surface, $A$, and the 
volume of the vesicle, $V$, are kept constant. The first constraint follows
from the large elastic compression modulus; the energy scale associated with 
this is much greater than that associated with the curvature 
elasticity and this multiplies a term which fixes $A$ \cite{sei97}. Similarly, 
the energy scale involving the osmotic pressure difference is so large, 
compared with the curvature energy, that the vesicle volume is effectively 
fixed. The shapes with the lowest bending energy are usually investigated at 
different values of a scaled spontaneous curvature and reduced volume. A 
``phase-diagram'' of these minimal shapes in these two variables has quite a 
complicated structure, with ``phases'' of prolate ellipsoids, dumb-bell and 
pear shapes appearing, amongst others \cite{sei97}.

Our intention here is to provide a means of linking these snapshots of the 
vesicle shape. To do this we need a dynamics which gives us a rule to move 
from one shape to another. We will be chiefly concerned with setting up the 
correct dynamical description of this system and so we will limit our 
attention to the growth of spherical vesicles and the transition from a 
spherical to an ellipsoidal shape. Our treatment differs from the previous 
studies of this problem\,\cite{boz04,boz07,mac07a,mac07b} in ways that we 
describe in detail in Section \ref{dyn_des}. However all these approaches 
have the common features that they use macroscopic variables and assume 
that the growth is sufficiently slow that it may be described within the 
formalism of linear irreversible thermodynamics.

The outline of the paper is as follows. In Section \ref{dyn_des} we introduce 
the formalism that will be used in the investigation and compare our approach 
to those used previously. In Section \ref{bifurcate} we carry out the 
analysis of the growth and loss of stability of the spherical vesicle in
the simplest case of a purely aqueous environment and in Section \ref{bif_sol}
we show how this generalizes when a solute is present. In 
Section \ref{conclude} we summarize our results and discuss them in the light 
of previous work and the model assumptions. There is a mathematical appendix 
which gives the technical details relating to Sections \ref{bifurcate} 
and \ref{bif_sol}.

\section{Dynamical Description}
\label{dyn_des}
In the spontaneous curvature model the membrane is a two-dimensional 
surface, $S$, which separates an inner region, $I$, from the environment, $E$. 
The outer region could be a purely aqueous environment, or it could also 
contain a solute, with both the water and solute molecules being able to 
permeate through the membrane. 

The surface is a purely geometric construction; it contains no matter and 
simply has a bending energy associated with it. If its shape is known, then 
this bending energy only depends on the volume, $V$, it encloses:
\begin{equation}
E(V) = \frac{\kappa}{2} \oint_{S} \left( C_{1}+C_{2}-C_{0} \right)^{2}\,dA\,,
\label{bend_ener}
\end{equation}
where $C_1$ and $C_2$ are the principal radii of curvatures of the surface. In
common with the other studies of this system\,\cite{boz04,boz07,mac07a,mac07b} 
the deviations from equilibrium will be taken to be sufficiently small that the
thermodynamic relation $TdS=dE+P_i dV+\sum_{\alpha} \mu_{\alpha}dc_{\alpha}$
can be used. Here $P_i$ is the pressure of the fluid inside the membrane, 
$\mu_{\alpha}$ the chemical potential of chemical $\alpha$ which has 
concentration $c_{\alpha}$. The only contribution from the membrane is a term 
$dE = (\partial E/\partial V)dV$, so changing the pressure inside the vesicle 
from $P_{i}$ to $(P_{i})_{\rm eff} = P_{i} + (\partial E/\partial V)$. 
Therefore, as long as we replace the internal pressure by this effective 
pressure, then we may ignore the membrane from a thermodynamic point of view, 
and simply treat it as a boundary which separates the inside of the vesicle 
from the environment.

\subsection{Purely aqueous environment}
\label{water}
The thermodynamic analysis of transport through a membrane which is simply
a geometric transition region between two homogeneous regions was carried
out by Kedem and Katchalsky fifty years ago\,\cite{ked58,ked63}. Suppose,
to begin with, that there is no solute present, so that only the flow of
water through the membrane need be considered. Then the usual assumption
of the thermodynamics of irreversible processes, that the processes under
consideration are sufficiently slow to give a linear relation between the
fluxes and the forces\,\cite{deG84}, lead to the relation\,\cite{ked58,ked63}
\begin{equation}
J_{w} = L_{p} \Delta P\,.
\label{F_F_1}
\end{equation}
Here $J_{w}$ is the flux of water from the environment to the interior,
$L_{p}$ is the hydraulic conductivity of the membrane and $\Delta P$ is
the difference between the exterior pressure, $P_e$, and that of the interior.
However, as discussed above, to include the contribution coming from the 
curvature of the membrane we need to replace $\Delta P$ by the effective 
pressure difference given by
\begin{equation}
(\Delta P)_{\rm eff} \equiv P_{e} - (P_{i})_{\rm eff} = P_{e} - P_{i} - 
\left( \frac{\partial E}{\partial V} \right) \equiv \Delta P -
\left( \frac{\partial E}{\partial V} \right).
\label{Delta_P}
\end{equation}

These results may now be brought together. The vesicle is assumed to increase
its surface area due to extra lipid molecules being added to the surface. 
This in turn will change the pressure in the interior, and so change 
$(\Delta P)_{\rm eff}$ and give rise to a flux of water through the membrane. 
The rate of increase of the volume of the vesicle will by given by
\begin{equation}
\frac{dV}{dt} = A J_{w} = L_{p} A \left[ \Delta P - 
\left( \frac{\partial E}{\partial V} \right) \right]\,,
\label{dV_by_dt}
\end{equation}
using Eqs.~(\ref{F_F_1}) and (\ref{Delta_P}). 

We now assume a growth law for the surface area, that is, the rate at which 
components are incorporated into the vesicle membrane. The simplest, and also
the most plausible, is that this is proportional to the surface 
area\,\cite{boz04}:
\begin{equation}
\frac{dA}{dt} = \lambda A\ \ \Rightarrow \ \ A(t) = A(0) e^{\lambda t}\,.
\label{growth_law}
\end{equation}
The analysis below can be carried out with other growth laws. At a more
fundamental level we would expect the correct form to emerge from the chemical
reactions underlying this process. It is convenient to define a reduced volume
by 
\begin{equation}
v = \frac{6\pi^{1/2} V}{A^{3/2}}\,,
\label{red_vol}
\end{equation}
so that $v=1$ for a sphere and $v<1$ for all other shapes. Then 
\begin{eqnarray}
\frac{dv}{dt} &=& - \frac{3}{2} \lambda v + \frac{6 \pi^{1/2}}{A^{3/2}} 
\frac{dV}{dt} \nonumber \\
&=& - \frac{3}{2} \lambda v + \frac{6 \pi^{1/2}}{A^{1/2}} L_{p} 
\left[ \Delta P - \left( \frac{\partial E}{\partial V} \right) \right]\,,
\label{v_eqn_water}
\end{eqnarray}
where we have used Eq.~(\ref{dV_by_dt}).

If the spontaneous curvature, $C_0$, was absent in the definition of the 
bending energy Eq.~(\ref{bend_ener}), then the energy would be scale invariant:
a typical length scale associated with the vesicle could be changed by an
arbitrary scaling factor and $E$ would remain unchanged. However, the inclusion
of $C_0$, which has dimensions of inverse length, introduces a scale into the
problem. Suppose that $R$ is the typical scale factor associated with the 
vesicle, then scaling the coordinates in Eq.~(\ref{bend_ener}) by this factor,
the typical length scale associated with the vesicle is unity. After rescaling
we denote the principal radii of curvatures by $k_{a}=C_{a}R$, $a=1,2$.
Similarly, a dimensionless spontaneous curvature, $k_{0} = C_{0}R$, may be 
introduced\,\cite{sei97}. If the vesicle is spherical, the typical length 
scale can be taken to be the radius, and $C_{1}=C_{2}=1/R$. Then 
\begin{equation}
E = 2 \pi \kappa \left( 2 - C_{0}R \right)^{2}\ \ \Rightarrow \ \
\frac{\partial E}{\partial V} = \frac{C_{0} \kappa}{R^2} \left(C_{0}R -
2 \right)\,.
\label{E_sphere}
\end{equation}

\subsection{Including a Solute}
\label{solute}
The formalism we have discussed in Section \ref{water} can be generalized to
include a solute. There will now be a flux of solute, $J_s$, in addition to
the flux of water $J_w$, and they will be linearly related to the 
thermodynamic driving forces which now includes the difference in the osmotic 
pressure of the solute across the membrane, $\Delta \Pi_{s}$, as well as 
$\Delta P$\,\cite{ked58,ked63}. The constants multiplying the forces in the 
linear relations are Onsager coefficients, which will be symmetric in the 
usual way\,\cite{deG84}. In fact we will use the linear relations involving a 
slightly different linear combination of variables, corresponding to the
``second set of practical phenomenological equations'' of Kedem and
Katchalsky\,\cite{ked63}:
\begin{eqnarray}
\label{F_F_2}
J_{v} &=& L_{p} \left( \Delta P - \sigma \Delta \Pi_{s} \right)\,, \\
J_{s} &=& \overline{c} \left( 1-\sigma \right) J_{v} + 
\omega \Delta \Pi_{s}\,,
\label{F_F_3}
\end{eqnarray}
where $\sigma$ is the reflection coefficient, $\overline{c}$ is the mean
concentration of the solute and $\omega$ is the solute permeability. The
flux $J_v$ is a linear combination of $J_w$ and $J_s$, namely 
$J_{v}=J_{w} \overline{V}_{w} + J_{s} \overline{V}_{s}$ where
$\overline{V}_{w}$ and $\overline{V}_{s}$ are the partial molar volumes of 
water and solute respectively. If we assume an ideal solute, then 
$\Delta \Pi_{s}=k_{B} T\Delta c$\,\cite{hil86}, where $k_{B}$ is Boltzmann's 
constant and $\Delta c$ the difference in concentrations across the membrane. 
This gives the more useful form
\begin{eqnarray}
\label{F_F_4}
J_{v} &=& L_{p} \left( \Delta P - \sigma k_{B} T \Delta c \right)\,, \\
J_{s} &=& \overline{c} \left( 1-\sigma \right) J_{v} + 
\omega k_{B} T \Delta c\,,
\label{F_F_5}
\end{eqnarray}
with $\Delta c = c_{e}-c_{i}$. 

In this case the total volume flow per unit area of the membrane is $J_v$,
and so Eq.~(\ref{dV_by_dt}) is replaced by 
\begin{equation}
\frac{dV}{dt} = A J_{v} = L_{p} A \left[ \Delta P - 
\left( \frac{\partial E}{\partial V} \right) - 
\sigma k_{B} T \Delta c \right]\,,
\label{dV_by_dt_sol}
\end{equation}
where, once again, we have replaced $\Delta P$ by $(\Delta P)_{\rm eff}$ to
account for the effect of the membrane curvature. Similarly, if $N$ is the 
number of molecules of the solute in the interior, then
\begin{equation}
\frac{dN}{dt} = A J_{s} = A \left[ \overline{c} \left( 1-\sigma \right) J_{v} 
+ \omega R T \Delta c \right]\,.
\label{dN_by_dt}
\end{equation}
This last result may be written in a number of different ways using the
relation $N=c_{i}V$.

\subsection{Comparisons with Previous Work}
\label{compare}
The analysis of the growth of vesicles presented in the next section will
start from Eqs.~(\ref{dV_by_dt_sol}) and (\ref{dN_by_dt}). However, we will
end this Section by discussing how these two equations differ from those
considered by previous workers investigating this problem. In 
Ref.~\cite{boz04} it was assumed that no solute was present, so the relevant 
discussion is that presented in Section \ref{water}. The equation which was 
used was not Eq.~(\ref{dV_by_dt}), however, since the term involving the 
bending energy was introduced in a very different way: the pressure difference
in Eq.~(\ref{F_F_1}) was simply set equal to $(\partial E/\partial V)$, 
resulting in the pressure difference being completely absent from 
Eq.~(\ref{dV_by_dt}). We believe our approach to be the correct way of 
proceeding. The method of Ref.~\cite{boz04} is extended to include a solute
in Ref.~\cite{boz07}. A further point of disagreement with our treatment is
that the reflection coefficient, $\sigma$, is set equal to unity. However 
this is only true if the membrane is impermeable to the solute, in which
case $\omega$ should equal zero too. The simultaneous use of 
Eq.~(\ref{dV_by_dt_sol}) and Eq.~({\ref{dN_by_dt}) when $\sigma=1$ was 
already argued against in the original paper of Kedem and 
Katchalsky\,\cite{ked58}. 

The work reported in Refs.~\cite{mac07a} and \cite{mac07b} has a different
philosophy; there it is assumed that the instability by which the division 
process begins --- the sphere becomes unstable to an ellipsoid --- is a Turing 
instability. Thus these authors introduce spatial effects, and the boundary 
of a two-dimensional vesicle is defined on a lattice. The ``total pressure''
involving the sum of the hydrostatic pressure difference, the osmotic 
pressure difference and the term coming from the surface energy are all
included. However, since this is a two-dimensional vesicle, the form of the
bending energy is different, and it is not clear to us why the surface tension 
is included with such a large modulus. Thermodynamic relations similar to 
Eqs.~(\ref{dV_by_dt_sol}) and (\ref{dN_by_dt}) are used, but apparently not 
to directly describe the change in vesicle shape. It is also not so clear to 
us what role the two metabolic centers that these authors introduce have in 
giving an initial anisotropy to the vesicle and how crucial they are in 
initiating the symmetry-breaking instability. Clearly subsequent divisions 
will produce vesicles without these centers and therefore the mechanism will 
have to still work in their absence.

One of the goals of the present work is to clarify the various assumptions 
made and systematize the methodology that is used to study vesicle growth and 
division.

\section{The first bifurcation}
\label{bifurcate}
In the purely static analysis of the model defined by Eq.~(\ref{bend_ener}),
it is known that when the sphere becomes unstable, the stable shape which
replaces it is the ellipsoid\,\cite{sei91}. As an initial application of the 
formalism of Section \ref{dyn_des}, we will investigate this instability from 
a dynamical viewpoint when no solute is present.

The axisymmetric ellipsoid will be parametrized by expressing the Cartesian 
coordinates as
\begin{eqnarray}
x &=& a \sin \theta \cos \phi\,, \nonumber \\
y &=& a \sin \theta \sin \phi\,, \nonumber \\
x &=& c \cos \theta\,,
\label{coords}
\end{eqnarray}
where $0 \leq \phi < 2\pi$, $-\pi/2 < \theta < \pi/2$ and where $a$ and $c$ 
are constants. For a sphere $a=c\equiv R$, the radius. If the ellipsoid 
only differs in shape from the sphere very slightly, then $a$ and $c$ may
be expressed as
\begin{equation}
a = R \left( 1 + a_{1} \epsilon \right)\,, \ \ c = R \left( 1 + c_{1} 
\epsilon \right)\,,
\label{a_and_c}
\end{equation}
where $\epsilon$ is a small quantity and $a_1$ and $c_1$ are numbers which 
characterize the shape of the ellipsoid; if $a_{1} > c_{1}$ it is oblate
and if $a_{1} < c_{1}$ it is prolate. 

Using standard results\,\cite{bey87} and (\ref{a_and_c}), it is 
straightforward to calculate the surface area and volume of the ellipsoid for 
small $\epsilon$. The details are given in the Appendix where it is shown that
\begin{eqnarray}
A &=& 4\pi R^{2} \left[ 1 + \frac{2}{3} \left( 2a_{1} + c_{1} \right)\epsilon
+ {\cal O} \left( \epsilon^{2} \right) \right]\,, \nonumber \\
V &=& \frac{4}{3} \pi R^{3} \left[ 1 + \left( 2a_{1} + c_{1} \right)\epsilon
+ {\cal O} \left( \epsilon^{2} \right) \right]\,.
\label{A_and_V}
\end{eqnarray}
From these expressions we see immediately that the reduced volume, $v$, 
defined by Eq.~(\ref{red_vol}) is $1 + {\cal O} (\epsilon^2)$, and so we
have to go to next order in Eq.~(\ref{A_and_V}) to find the deviation of the 
reduced volume from the value $1$ which it has when the vesicle is spherical.
From Eqs.~(\ref{SA_3}) and (\ref{Vol_2}) we see that
\begin{equation}
v = 1 - \frac{4}{15} \left( a_{1}-c_{1} \right)^{2} \epsilon^{2} 
+ {\cal O} \left( \epsilon^{3} \right)\,,
\label{v_to_second}
\end{equation}
with $v<1$ for all cases except the sphere ($a_{1}=c_{1}$) as required. The
bending energy is also straightforward to calculate, but much more tedious.
From it we can determine $\partial E/\partial V$, as described in the 
Appendix. 
 
Substituting the expressions for $v$ and $\partial E/\partial V$ into 
Eq.~(\ref{v_eqn_water}) we find
\begin{equation}
\frac{8}{15\,\ln 2} \left( a_{1}-c_{1} \right)^{2} \epsilon 
\frac{d\epsilon}{d\tau} = F_{0} + F_{1} \epsilon + F_{2} \epsilon^{2} 
+ {\cal O} \left( \epsilon^{3} \right)\,,
\label{water_bal_eqn}
\end{equation}
where the $F_j$ ($j=0,1,2$) are functions of $R, \Delta P, C_{0}$ and 
$\eta (\equiv L_{p} \kappa C^{4}_{0} \ln 2/\lambda)$ and are given explicitly
in Eq.~(\ref{Fs}). We have also introduced a scaled time $\tau = t/T_{d}$, 
where $T_{d}=\ln 2/\lambda$ is the time taken for the surface area to double in
value. For Eq.~(\ref{water_bal_eqn}) to be consistent as $\epsilon \to 0$ we 
require $F_{0}=0$, which gives $\Delta P$ in terms of $R, C_{0}$ and $\eta$:
\begin{equation}
\frac{\Delta P}{\kappa} = \frac{C_{0}^{4} R \ln 2}{2 \eta} 
+ \frac{C_{0}^{2}}{R} - \frac{2 C_0}{R^2}\,.
\label{DeltaP_water}
\end{equation}
 
Since Eq.~(\ref{DeltaP_water}) holds as $\epsilon \to 0$, it is true for the 
sphere and could have been obtained more directly from the bending energy of 
a spherical vesicle given in Eq.~(\ref{E_sphere}). Then, since for a sphere  
\begin{displaymath}
\frac{dV}{dt} = \frac{R}{2} \frac{dA}{dt} = \frac{\lambda R}{2} A\,,
\end{displaymath}
we see from Eq.~(\ref{dV_by_dt}) that 
\begin{equation}
\frac{\lambda R}{2} = L_{p} \left[ \Delta P - 
\frac{\kappa C_{0} R (C_{0} R - 2)}{R^3} \right]\,,
\label{zeroth_directly}
\end{equation}
which agrees with Eq.~(\ref{DeltaP_water}). The quantity $\Delta P$ is the 
pressure difference required for the vesicle to remain a sphere while growing 
at a steady rate given by $\lambda$.
 
Setting $F_{0}=0$ in Eq.~(\ref{water_bal_eqn}), one sees that the two sides
of the equation are not of the same order as $\epsilon \to 0$ unless 
$F_{1}=0$. From the explicit form for this function given in the Appendix,
setting $\Delta P$ equal to the value given in Eq.~(\ref{DeltaP_water}) gives
the result displayed in Eq.~(\ref{F_1}). We see that, apart from making a 
particular choice for $R$ in terms of $C_0$ and $\eta$, we can only make 
$F_{1}$ equal to zero by taking $(2a_{1}+c_{1})=0$. This simply amounts to a 
particular choice of ellipsoid shape. It is, in some sense, the most 
symmetrical choice and consists of changes in the direction of the two 
symmetric axes being half of that in the third direction (and having the 
opposite sign). 

Finally, setting both $F_0$ and $F_1$ equal to zero, which implies the 
choice Eq.~(\ref{DeltaP_water}) and $c_{1}=-2a_{1}$, gives the expression
(\ref{F_2_final}) for $F_2$ and leads to 
\begin{equation}
\frac{d\epsilon}{d\tau} = \left( \frac{13 \eta}{8 C^2_0 R^2} 
- \frac{9 \eta}{2 C^3_0 R^3} - \frac{13 \ln 2}{16} \right) \epsilon
+ {\cal O} \left( \epsilon^{2} \right)\,.
\label{stab_cond_water}
\end{equation}
If the term on the right-hand side of this equation is positive, then the
sphere will be unstable to a transformation into an ellipsoid. This is the 
case if $2\eta (13 C_{0}R - 36) > 13 C_{0}^3 R^3 \ln 2$, which can never be
satisfied if $C_{0}R < 36/13$. Similarly when viewed as an inequality which 
is cubic in $C_{0} R$, one finds that it cannot be satisfied for any real 
positive $C_{0} R$ if $\eta < \eta_{\rm min}$ where
\begin{equation}
\eta_{\rm min} = \frac{3^{7}\,2 \ln 2}{13^2} \approx 17.94\,.
\label{eta_min}
\end{equation}
For $C_{0}R > 36/13$ the condition for the sphere to be unstable may be written
as
\begin{equation} 
\eta > \frac{1}{2} \left( \frac{13 C_0^3 R^3 \ln 2 }{13 C_0 R - 36} \right)\,. 
\label{eta_boundary_water}
\end{equation}
The region of instability is shown in Fig.~\ref{fig2}. For values of 
$\eta>\eta_{\rm min}$ a dynamical transition can eventually occur, the vesicle 
turning into an ellipsoid. We are only interested in the first transition 
which is encountered as $R$ increases, which occurs for values of $C_{0}R$ 
greater than $36/13$, but less than the value that corresponds to  
$\eta=\eta_{\rm min}$. This latter condition gives the result
$C_{0} R = 54/13$. Therefore the critical radius, $R_{c}$, at which the
transition occurs, lies in the range $36/13< C_{0} R_{c} < 54/13$. It is 
interesting to note the fundamental role that the phenomenological factor 
$C_0$ has in determining the critical radius. 

\begin{figure}[htbp]
\centering
\vspace*{2.5em}
\includegraphics[width=7cm]{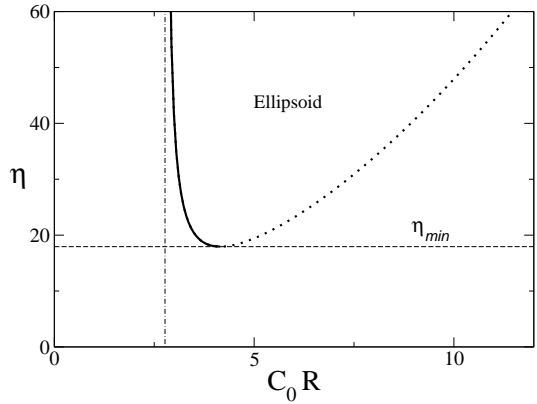}
\caption{The transition line defining the region of instability in the 
parameter plane ($C_0 R$, $\eta$) is depicted for a purely aqueous environment.
There is no transition below the horizontal dashed line (which represents 
$\eta_{\rm min}=17.94$) and to the left of the vertical dashed-dotted line 
(which represents $C_{0}R = 36/13$). In the remaining portion of the figure, 
the thick solid line represents the transition from the region on the left 
where the spherical configuration is dynamically favored, to the region on the 
right, where the ellipsoid configuration is dynamically favored. The dotted 
line is an unphysical solution of the cubic equation (\ref{cubic}).}
\label{fig2}
\end{figure}

So, in summary, we suppose that initially the vesicle is a sphere of radius 
$R(0)$. It then grows according to Eq.~(\ref{growth_law}), that is, the
radius increases according to $R(t)=R(0)e^{\lambda t/2}$. The pressure
difference between the interior of the vesicle and the exterior during this 
growth phase may be found from Eq.~(\ref{DeltaP_water}) to be
\begin{equation}
\frac{\Delta P}{\kappa} = \frac{C^4_0 R(0) \ln 2}{2\eta} e^{\lambda t/2}
+ \frac{C^2_0}{R(0)} e^{-\lambda t/2} - \frac{2C_0}{R(0)^2} e^{-\lambda t}\,.
\label{Deltap_time}
\end{equation}
The growth phase continues until the vesicle has achieved a radius of $R_c$, 
which is the smallest real positive root of the cubic equation found by 
setting $F_{2}$ equal to zero:
\begin{equation} 
13 C_0^3 R_{c}^3 \ln 2 - 26 \eta C_0 R_{c} + 72 \eta = 0\,.
\label{cubic}
\end{equation}
As discussed above there are no such roots for $\eta < \eta_{\rm min}$, given 
by Eq.~(\ref{eta_min}), and for $\eta > \eta_{\rm min}$, $C_{0} R_{c}$ has to 
lie in the narrow range $[2.77, 4.15]$. The critical radius is reached at a 
time
\begin{equation}
t_{c} = \frac{2}{\lambda} \ln \left( \frac{R_{c}}{R(0)} \right) \ \ 
{\rm or} \ \ \tau_{c} = \frac{\ln \left( R_{c}/R(0) \right)^{2}}{\ln 2}\,.
\label{crit_time}
\end{equation}
At this time the spherical shape becomes unstable and the vesicle takes on an 
ellipsoidal shape.

\section{The bifurcation with a solute present}
\label{bif_sol}
In Section \ref{dyn_des} we developed the formalism for the situation where
a solute was present, but for simplicity the analysis of 
Section \ref{bifurcate} assumed that the solute was absent. In this Section 
we will repeat the analysis of Section \ref{bifurcate} with the solute 
included. 

The extra term which appears in Eq.~(\ref{dV_by_dt_sol}) which changes the
position of the instability is $-\sigma k_{B} T \Delta c$. To calculate it
we need to determine $\Delta c$. This can be found from the other equation we
introduced, Eq.~(\ref{dN_by_dt}), however an integral over time has to be
performed. To see this, we use Eq.~(\ref{dV_by_dt_sol}) to write 
Eq.~(\ref{dN_by_dt}) as
\begin{equation}
\frac{d \left( c_{i} V \right) }{dt} = \overline{c} \left( 1-\sigma \right) 
\frac{dV}{dt} + A\,\omega k_{B} T \Delta c\,,
\label{dN_by_dt_2}
\end{equation}
where $\overline{c} = \Delta c/\ln(c_{e}/c_{i})$\,\cite{ked58}. If we know 
$A$ and $V$ as functions of $t$, then we can in principle solve this
differential equation for $c_{i}(t)$ and so find $\Delta c(t)$, which can then
be substituted into Eq.~(\ref{dV_by_dt_sol}).

To illustrate these basic ideas, we will consider the special case where the
membrane is impermeable to solute molecules, so that the number of solute 
molecules, $N$, is constant. However $c_{i}(t)$ does change with time, due to
the fact that the volume, $V$, increases with time, and this has a non-trivial
effect on the instability analysis. A membrane impermeable to solute molecules
is defined by $\sigma=1$ and $\omega=0$. It follows directly from 
Eq.~(\ref{dN_by_dt_2}) that $c_{i} V$ is a constant, and so 
\begin{equation}
c_{i} (t) = \frac{c_{i} (0) V(0)}{V(t)}\,.
\label{c_i_growth}
\end{equation}
Then the term $-k_{B} T \Delta c$ which appears in Eq.~(\ref{dV_by_dt_sol}) 
gives rise to an additional term on the right-hand side of 
Eq.~(\ref{v_eqn_water}) which equals
\begin{equation}
- \frac{6\pi^{1/2}}{A^{1/2} V} L_{p} k_{B} T \left\{ c_{e} V(t)
- c_{i} (0) V(0) \right\}\,.
\label{extra_term}
\end{equation}
Using the expressions (\ref{A_and_V}), this becomes
\begin{eqnarray}
&-& \frac{3L_{p} k_{B} T}{R^4} 
\left\{ 1 - \frac{4}{3} \left( 2a_{1} + c_{1} \right)\epsilon
 + {\cal O} \left( \epsilon^{2} \right) \right\} \nonumber \\
&\times&  \left\{ c_{e} R^{3} 
\left[ 1 + \left( 2 a_{1} + c_{1} \right)\epsilon 
+ {\cal O} \left( \epsilon^{2} \right) \right] - c_{i} (0) R^{3} (0)\right\}\,,
\nonumber \\
\label{to_first_order}
\end{eqnarray}
to first order in $\epsilon$, since initially the vesicle is a sphere.

To see how these changes affect Eq.~(\ref{water_bal_eqn}) let us write it as
\begin{equation}
\frac{8}{15\,\ln 2} \left( a_{1}-c_{1} \right)^{2} \epsilon 
 \frac{d\epsilon}{d\tau} = G_{0} + G_{1} \epsilon + G_{2} \epsilon^{2} 
+ {\cal O} \left( \epsilon^{3} \right)\,,
\label{solute_bal_eqn}
\end{equation}
where the $G_j$ ($j=0,1$) are related to the $F_j$ as follows:
\begin{eqnarray}
\label{G_0}
G_{0} &=& F_{0} - \frac{3\eta \Phi \left[ 3 \Gamma - 4\pi R^{3} 
c_{e} \right] }{4 \pi C^{4}_{0} R^{4} \ln 2} \\
G_{1} &=& F_{1} + \frac{\eta \Phi \left[ 3 \Gamma - \pi R^{3} 
c_{e} \right] }{\pi C^{4}_{0} R^{4} \ln 2}\,,
\label{G_1}
\end{eqnarray}
where $\Gamma = c_{i}(0) V(0)$ and $\Phi = \sigma k_{B} T$. The analogous 
result for $G_2$ is given by Eq.~(\ref{G_2}) in the Appendix. Following the
same line of argument as in Section \ref{bifurcate} we require $G_{0}$ and
$G_{1}$ to be zero in order for the stability analysis to be applicable. The 
first condition gives an expression for the pressure difference across the 
membrane:
\begin{eqnarray}
\frac{\Delta P}{\kappa} &=& \frac{C_{0}^{4} R \ln 2}{2 \eta} 
+ \frac{C_{0}^{2}}{R} - \frac{2 C_0}{R^2} \nonumber \\
&+& \frac{\sigma k_{B} T}{R^3} \left( R^{3} c_{e} - R^{3}(0)c_{i}(0) \right)\,,
\label{DeltaP_solute}
\end{eqnarray}
which shows the additional terms that are added to Eq.~(\ref{DeltaP_water}) 
when a solute is present. The second condition, $G_{1}=0$ again implies that
$(2a_{1}+c_{1})=0$ and so the addition of the solute does not change the 
shape of the ellipsoid for which the stability analysis applies.

If we now use the conditions found by implementing $G_{0}=0$ and $G_{1}=0$ in 
Eq.~(\ref{G_2}) and substituting this into Eq.~(\ref{solute_bal_eqn}) we find
\begin{eqnarray}
\frac{d\epsilon}{d\tau} &=& \left( \frac{13 \eta}{8 C^2_0 R^2} 
- \frac{9 \eta}{2 C^3_0 R^3} - \frac{13 \ln 2}{16} \right. \nonumber \\ 
&-& \left. \frac{45\Gamma\Phi\eta}{32\pi C^{4}_{0} R^{4}}\right)\epsilon
+ {\cal O} \left( \epsilon^{2} \right)\,,
\label{stab_cond_solute}
\end{eqnarray}
which gives the required modification of Eq.~(\ref{stab_cond_water}). A similar
analysis to that given in Section \ref{bifurcate} shows that if
$4\pi C_{0} R (13 C_{0} R - 36) < 45 \Gamma \Phi$, then the sphere is always 
stable, no matter what the value of $\eta$. This gives the minimum value of
the radius, $R$, which can lead to an instability (corresponding to an 
infinite value for $\eta$) to be
\begin{equation}
C_{0} R_{\rm min} = \frac{18}{13} \left[ 1 + \sqrt{ 1 + 
\frac{65 \Gamma \Phi}{144 \pi}} \right]\,.
\label{R_min}
\end{equation}
If $R > R_{\rm min}$, the spherical shape is unstable if
\begin{equation} 
\eta > \left( \frac{26 \pi C_0^4 R^4 \ln 2 }
{4\pi C_{0} R \left( 13 C_0 R - 36\right) - 45\Gamma \Phi } \right)\,. 
\label{eta_boundary_solute}
\end{equation}
 
Figure~\ref{fig3} depicts the domain of instability in the parameter plane
$(C_0 R, \eta)$. Different curves refer to different choices of the quantity 
$\Gamma \Phi$, and allow the qualitative inspection of the modification induced
by the presence of a solute. The condition for a double root of the quartic 
equation for $C_{0}R$ gives, as in Section \ref{bifurcate}, the maximum 
value of $R$ which can lead to an instability:
\begin{equation}
C_{0} R_{\rm max} = \frac{27}{13} \left[ 1 + \sqrt{ 1 + 
\frac{65 \Gamma \Phi}{162 \pi}} \right]\,.
\label{R_max}
\end{equation}
with the corresponding value of $\eta$ being $\eta_{\rm min}$. This can be 
determined as a function of $\Gamma \Phi$ by substituting the expression
(\ref{R_max}) for $C_{0} R_{\rm max}$ back into the quartic equation. The 
resulting function is shown as an inset in Fig.~\ref{fig3}. This increases as 
$\Gamma \Phi$ increases (larger solute concentration and/or higher 
temperature). However, the ratio of $R_{\rm max}/R_{\rm min}$ remains 
essentially unchanged at about $1.5$ for all values of $\Gamma \Phi$. This
means that the range of values of the radius at which the spherical shape 
becomes unstable remains quite small.

\begin{figure}[htbp]
\centering
\vspace*{2.5em}
\includegraphics[width=7cm]{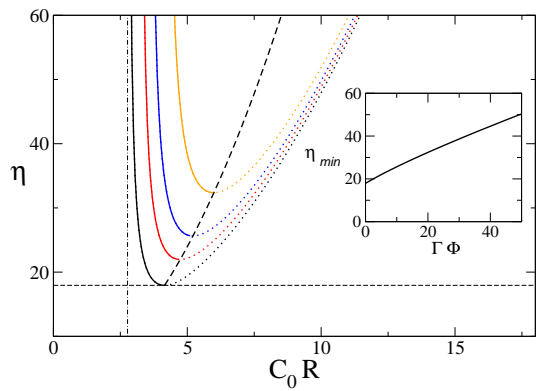}
\caption{(Color online) The region of instability is depicted for the case 
when a solute is included in a similar manner to that shown in Fig.~\ref{fig2} 
when a solute is not present. Different curves refer to distinct values of 
the product $\Gamma \Phi$ (= 0, 5, 10, 20). The dotted line which passes 
through the minima of these curves shows the maximum value of $C_{0} R$ for
which a transition can occur, and is given by Eq.~(\ref{R_max}). The
corresponding value of $\eta$, $\eta_{\rm min}$, is plotted in the inset as a 
function of the quantity $\Gamma \Phi$.}
\label{fig3}
\end{figure}

\section{Conclusion}
\label{conclude}

Despite the great interest in the growth, change in shape and division of 
vesicles, very little is known about the nature of the processes that govern 
them. Even fundamental questions about the typical vesicle radii at the 
various stages or the hydrostatic or osmotic pressure differences between 
the exterior and interior, are still largely open.  The main obstacle to 
achieving a greater understanding is the difficulty in carrying out 
experiments. Even qualitatively, a consistent picture is hard to achieve,
and so a theoretical description which would help with the interpretation of 
experimental results would be very welcome. In this paper we have proposed 
such a description, taking extra care to correctly incorporate the energy 
associated with the curvature of the membrane in the thermodynamic description.
We have concentrated on establishing the formalism and demonstrating it on 
the initial stages of the growth and on the first bifurcation to an ellipsoid. 
The subsequent time evolution of the vesicle, leading to division, can now be 
investigated, but this can only be carried out numerically and we leave it 
for the future.

The vesicle growth is ultimately caused by the incorporation of lipids from 
the environment into the vesicle wall. This will increase the area, which in 
turn will lead to a change in the internal hydrostatic pressure and so initiate
a flow of fluid through the membrane. This will happen extremely 
slowly --- at a rate governed by the parameter $\lambda$, and so in practice 
the vesicle will never become flaccid and will maintain a spherical shape 
in the initial stages of the evolution. It is the quasi-static nature of
the expansion that allows us to use the formalism of linear irreversible 
thermodynamics\,\cite{deG84}.  This predicts, among other things, that the 
radius of the vesicle will grow according to 
$R(t)=R(0) e^{\lambda t/2} = 2^{t/(2T_{d})}$ using $\lambda=\ln 2/T_d$.
However this exponential growth gets cut off at a radius, $R_c$, given by the 
smallest real solution of Eq.~(\ref{cubic}) at a critical time given by 
Eq.~(\ref{crit_time}). This critical radius falls into a remarkably narrow
range. The corresponding value of the hydrostatic pressure can then be 
determined.

It is interesting to compare these results with those found by the previous
studies discussed in Section \ref{compare}. In Refs.\,\cite{boz04,boz07}, the
term involving the bending energy on the right of Eq.~(\ref{dV_by_dt}) was 
effectively omitted and therefore the expression (\ref{DeltaP_water}) did not 
contain the final two terms, which originate from this term (see 
Eq.~(\ref{E_sphere})). It could be argued that it is a legitimate 
approximation, in the sense that for the parameters of interest, this term is 
negligible compared to the one that is retained. Since we have included both 
contributions, we are able to test this by calculating the ratio between the 
first term on the right-hand side of Eq.~(\ref{DeltaP_water}) and the sum of 
the second and third terms. We find that this ratio always lies between $0$ 
and $1$. That is, the terms omitted in Refs.\,\cite{boz04,boz07} are always 
more important than the one included. This ratio is actually zero at one 
extreme of the range of allowed values of $R$ ($C_{0}R=36/13$, which 
corresponds to an infinite value of the parameter $\eta$), and increases 
monotonically to $9/14$ when $C_{0}R$ takes on its greatest allowed value 
($54/13$) and $\eta$ takes on its smallest allowed value ($17.94$). It should 
also be remarked that in Ref.~\,\cite{boz04} the value of $\eta$ used is 
$1.85$, which is the value at which vesicle splitting gives rise to two equal 
vesicles, which are equal in size to the initial vesicle. Given the 
inconsistencies in the formulation we have just alluded to, it is doubtful 
that this value is correct. Further work is required to enable a comparison 
between the work reported in Refs.~\cite{mac07a} and \cite{mac07b} and our 
approach. As explained in Section \ref{compare}, the method used is very 
different: the vesicle is a closed curve in two dimensions, metabolic centers 
are present which induce the symmetry breaking and the results obtained are 
purely numerical.

These results, and others involving the nature of the initial bifurcation, 
provide an initial set of predictions which can be compared with future 
experiments. Of course, the model as it stands is quite simple and many 
features could be made more realistic. For instance, $C_0$ is almost certainly 
not a constant, and becomes non-zero during the course of the vesicle growth. 
Nevertheless, taking a non-zero $C_0$ induces a transition\,\cite{polm}. 
Another aspect that we have not included in the present treatment is a 
discussion of the effect of thermal fluctuations on the shape of the vesicle. 
At first sight it is not obvious if thermal fluctuations will have any 
appreciable effect. An order of magnitude calculation\,\cite{wor97} suggests 
that under most circumstances fluctuations will not need to be taken into 
account, however near an instability there may be a significant effect. Within 
the static picture, which has been the subject of by far the majority of papers
to date, the procedure for investigating the effect of thermal fluctuations 
is clear\,\cite{hel84,sei94,hei97,wor97,far03}. The shape is written as a sum 
of a stationary term plus a small fluctuation and the energy expanded to 
quadratic order in the fluctuation. Putting this into a Boltzmann factor, the 
Gaussian integrals may be performed. However there are a number of technical 
issues, such as the inclusion of constraints. There is also the question of 
the size of non-Gaussian fluctuations. Nevertheless, it is clear that thermal 
fluctuations can shift the position of an instability, or even change its 
nature completely. The analogous calculation carried out within the present 
framework would be even more complex, however it should definitely be addressed
in future work.  

Despite such shortcomings, and there are undoubtedly others, we believe that 
the model is sufficiently detailed to provide a reasonably good description 
of vesicle growth and hope that it will serve to clarify a number of aspects 
in this fascinating, and neglected, field.

\begin{acknowledgments}

We wish to thank Timoteo Carletti, Pier Luigi Luisi, Javier Macia, Peter 
Olmsted, Roberto Serra, Pasquale Stano, Sasa Svetina and Matthew Turner
for interesting discussions and correspondence. AJM wishes the thank the
EPSRC (UK) for financial support under grant GR/T11784/01.

\end{acknowledgments}

\appendix

\section{Surface area, volume and bending energy of an ellipsoid}
\label{append_ell}
In this Appendix we will collect together the results which are required to
calculate the reduced volume and the bending energy of an ellipsoid in Section
\ref{bifurcate}, together with the generalization in Section \ref{bif_sol}.

The ellipsoids we consider here are axisymmetric. The prolate version is
formed by rotating an ellipse with semi-minor axis $a$ and semi-major axis $c$ 
(i.e. $c>a$) about the major axis. It surface area is given by\,\cite{bey87}
\begin{equation}
A = 2\pi a^{2} + \frac{2\pi a c^{2}}{\sqrt{c^{2}-a^{2}}} \sin^{-1} 
\left[ \frac{\sqrt{c^{2}-a^{2}}}{c} \right]\,.
\label{SA_1}
\end{equation}
For an oblate ellipsoid, $c<a$, and\,\cite{bey87},
\begin{equation}
A = 2\pi a^{2} + \frac{\pi a c^{2}}{\sqrt{a^{2}-c^{2}}} 
\ln \left[ \frac{a + \sqrt{a^{2}-c^{2}}}{a - \sqrt{a^{2}-c^{2}}} \right]\,.
\label{SA_2}
\end{equation}
In both cases their volume is given by\,\cite{bey87}
\begin{equation}
V = \frac{4}{3}\pi a^{2}c\,.
\label{Vol_1}
\end{equation}

Substituting the parametrizations (\ref{a_and_c}) into the expressions
(\ref{SA_1})-(\ref{Vol_1}) one finds Eq.~(\ref{A_and_V}). As mentioned in the
text, this immediately implies that the reduced volume is 1 to this order,
and so we have to go to next order in $\epsilon$ to see some deviation from 
the result for a sphere. At next order
\begin{eqnarray}
A &=& 4\pi R^{2} \left[ 1 + \frac{2}{3} 
\left( 2a_{1} + c_{1} \right)\epsilon \right. \nonumber \\
&+& \left. \frac{1}{15} \left( 6a_{1}^{2} + c_{1}^{2} + 8a_{1} c_{1} \right)
\epsilon^2 + {\cal O} \left( \epsilon^{3} \right) \right]\,,
\label{SA_3}
\end{eqnarray}
and
\begin{eqnarray}
V &=& \frac{4}{3} \pi R^{3} \left[ 1 + \left( 2a_{1} 
+ c_{1} \right)\epsilon \right. \nonumber \\
&+& \left. \left( a^{2}_{1} + 2 a_{1} c_{1} \right) \epsilon^{2}
+ {\cal O} \left( \epsilon^{3} \right) \right]\,,
\label{Vol_2}
\end{eqnarray}
which together give the expression (\ref{v_to_second}) for $v$ to second 
order in $\epsilon$.

The other quantity we have to evaluate is the bending energy, $E$, given by
Eq.~(\ref{bend_ener}). This involves evaluating the two integrals 
\begin{equation}
J_{1} \equiv \oint_{S} H\,dA\,, \ \  J_{2} \equiv \oint_{S} H^{2}\,dA\,,
\label{defn_Js}
\end{equation}
where $H=(C_{1}+C_{2})/2$ is the mean curvature. For an axisymmetric ellipsoid 
this is given by\,\cite{bey87}
\begin{equation}
H = \frac{c\left[ 3a^{2}+c^{2}+\left( a^{2}-c^{2} \right) \cos 2\theta \right]}
{ \left[ a^{2}\cos^{2}\theta + c^{2}\sin^{2}\theta \right]^{3/2}}\,.
\label{H_ellipsoid}
\end{equation}
Evaluating the integrals in Eq.~(\ref{defn_Js}) using the result 
(\ref{H_ellipsoid}) yields
\begin{eqnarray}
J_{1} &=& 2\pi c + \frac{\pi a^{2}}{\sqrt{c^{2}-a^{2}}} 
\ln \left[ \frac{c + \sqrt{c^{2}-a^{2}}}{c - \sqrt{c^{2}-a^{2}}} \right]
\ \ (c>a)\,,\nonumber \\
J_{1} &=& 2\pi c + \frac{2\pi a^{2}}{\sqrt{a^{2}-c^{2}}} \sin^{-1} 
\left[ \frac{\sqrt{a^{2}-c^{2}}}{a}\right] \ \ (c<a)\,, \nonumber \\
\label{J_1s}
\end{eqnarray}
and
\begin{eqnarray}
J_{2} &=& \frac{\pi c^2}{a} \left\{ \frac{\sin^{-1} 
\left[ \frac{\sqrt{c^{2}-a^{2}}}{c} \right] }{\sqrt{c^{2}-a^{2}}} 
+ \frac{7a}{3c^2} - \frac{2 a^3}{3 c^4} \right\} \ \ (c>a)\,,\nonumber \\
J_{2} &=& \frac{\pi c^2}{a} \left\{ \frac{{\rm tanh}^{-1} 
\left[ \frac{\sqrt{a^{2}-c^{2}}}{a} \right] }{\sqrt{a^{2}-c^{2}}} 
+ \frac{7a}{3c^2} - \frac{2 a^3}{3 c^4} \right\} \ \ (c<a)\,. \nonumber \\
\label{J_2s}
\end{eqnarray}
Substituting the parametrizations (\ref{a_and_c}) into the expressions
(\ref{J_1s}) and (\ref{J_2s}), and using Eq.~(\ref{SA_3}), one obtains
\begin{eqnarray}
E &=& 2 \pi \kappa \left\{ \left(C_0 R - 2 \right)^2 + 
\frac{2}{3} \left(2 a_1 + c_1 \right) C_0 R \left(C_0 R - 2 \right) \epsilon 
\right. \nonumber \\
&+& \frac{1}{15} \left[c_1^2 \left(32-4 C_0 R + C^2_0 R^2 \right) + 
8 a_1 c_1 \left( -8 + C_0 R \right. \right. \nonumber \\
&+& \left. \left. \left. C_0^2 R^2 \right) + a_1^2 \left( 32 - 4 C_0 R 
+ 6 C_0^2 R^2 \right) \right] \epsilon^2 
+ {\cal O} \left( \epsilon^{3} \right) \right\} \,. \nonumber \\
\label{ener_second}
\end{eqnarray}
The aim is to calculate $\partial E/\partial V$ which we achieve by using
\begin{equation} 
\frac{\partial E}{\partial V} = \frac{\partial E}{\partial R} \frac{dR}{dV}\,.
\label{dE_over_dV}
\end{equation}
From Eq.~(\ref{ener_second}) one finds
\begin{eqnarray}
\frac{\partial E}{\partial R} &=& 4 \pi \kappa C_{0} 
\left\{ \left(C_0 R - 2 \right) + \frac{2}{3} \left( 2 a_1 + c_1 \right) 
\left(C_0 R - 1 \right) \epsilon \right. \nonumber \\
&+& \frac{1}{15} \left[ c_1^2 \left( C_0 R - 2 \right) + 
4 a_1 c_1 \left( 1 + 2 C_0 R \right) \right. \nonumber \\
&+& \left. \left. a_1^2 \left( 6 C_0 R - 2 \right) \right] \epsilon^2 
+ {\cal O} \left( \epsilon^{3} \right) \right\} \,,
\label{ener_derivative}
\end{eqnarray}
and from Eq.~(\ref{Vol_2})
\begin{eqnarray}
\frac{dV}{dR} &=& 4 \pi R^2 \left[ 1 + \left( 2a_{1} 
+ c_{1} \right)\epsilon \right. \nonumber \\
&+& \left. \left( a^{2}_{1} + 2 a_{1} c_{1} \right) \epsilon^{2}
+ {\cal O} \left( \epsilon^{3} \right) \right]\,,
\label{dV_by_dR}
\end{eqnarray}
which gives an expression for $\partial E/\partial V$. Substituting this into 
Eq.~(\ref{v_eqn_water}), and making use of the expression (\ref{v_to_second})
for $v$ we obtain Eq.~(\ref{water_bal_eqn}) with the $F_j$ ($j=0,1,2$) given 
by
\begin{eqnarray}
F_{0} &=& 3 \left( \frac{C_0^4 R^3 \ln 2 - 4 C_0 \eta + 2 C_0^2 \eta R 
- 2 \Delta \tilde{P} \eta R^2}{2 C_0^4 R^3 \ln 2} \right)\,, \nonumber \\
F_{1} &=& \left( 2 a_1 + c_1 \right) \eta \left( \frac{6 C_0 - 2 C_0^2 R  
+ \Delta \tilde{P} R^2}{C_0^4 R^3 \ln 2} \right)\,, \nonumber \\  
F_{2} &=& \frac{1}{15 C_0^4 R^3 \ln 2} \left[ 2 a_1 c_1 (-82 C_0 \eta + 
16 C_0^2 \eta R \right. \nonumber \\
&-& 9 \Delta \tilde{P} \eta R^2 + 6 C_0^4 R^3 \ln 2) - c_1^2 (98 C_0 \eta 
- 29 C_0^2 \eta R \nonumber \\
&+& 6 \Delta \tilde{P} \eta R^2 + 6 C_0^4 R^3 \ln 2) - a_1^2 (278 C_0 \eta 
\nonumber \\ 
&-& \left. 74 C_0^2 \eta R + 21 \Delta \tilde{P} \eta R^2 
+ 6 C_0^4 R^3 \ln 2) \right]\,,
\label{Fs}
\end{eqnarray}
where $\Delta \tilde{P} = \Delta P/\kappa$.
 
As explained in the main text, setting $F_{0}=0$ gives 
Eq.~(\ref{DeltaP_water}). Using this expression for $\Delta P$ then gives
\begin{equation}
F_{1} = \left( 2 a_1 + c_1 \right) \left( \frac{ C^{3}_{0} R^{3} \ln 2
+ 8\eta - 2\eta C_{0} R}{C_0^3 R^3 \ln 2} \right)\,.
\label{F_1}
\end{equation}
Apart from exceptional values of $R$, this vanishes only when 
$2a_{1}+c_{1}=0$. 

If we now set $c_{1}=-2a_{1}$ we find from Eq.~(\ref{Fs}) that
\begin{equation} 
F_{2} = 3 a_1^2 \left( \frac{14 C_0^2 \eta R - 38 C_0 \eta - \Delta \tilde{P}
\eta R^2 - 6 C_0^4 R^3 \ln 2}{5 C_0^4 R^3 \ln 2} \right)\,.
\label{F_2}
\end{equation}
Finally, setting $\Delta P$ to the value given in Eq.~(\ref{DeltaP_water}) 
one finds
\begin{equation} 
F_{2} = 3 a_1^2 \left( \frac{26 \eta C_0 R - 13 C_0^3 R^3 \ln 2 - 72 \eta}
{10 C_0^3 R^3 \ln 2} \right)\,,
\label{F_2_final}
\end{equation}
which leads to Eq.~(\ref{stab_cond_water}).

In Section \ref{bif_sol}, the calculation is carried out in the present of 
a solute. The equation which describes the instability is now 
Eq.~(\ref{solute_bal_eqn}) with $G_0$ and $G_1$ given by 
Eqs.~(\ref{G_0}) and (\ref{G_1}) respectively and $G_2$ given by
\begin{eqnarray}
G_{2} &=& F_{2} - \frac{\eta \Phi}{10\pi C^{4}_{0} R^{4} \ln 2} 
\left[ 4a_{1}c_{1} \left( 21 \Gamma - 3\pi R^{3} c_{e} \right) \right. 
\nonumber \\
&+& \left. c^{2}_{1} \left( 33 \Gamma - 4\pi R^{3} c_{e} \right)
+ 2a^{2}_{1} \left( 54 \Gamma - 7\pi R^{3} c_{e} \right) \right]\,,
\nonumber \\
\label{G_2}
\end{eqnarray}
where once again $\Gamma = c_{i}(0) V(0)$ and $\Phi = \sigma k_{B} T$. 


\end{document}